\documentstyle[epsf,PASJadd]{PASJ95} 
\def\tatehaba2{\rule[-1.5mm]{3mm}{5mm}}
\def\tatehaba3{\rule[1mm]{2mm}{0.5mm}}

\begin{document}


\title{Discovery of Non Thermal X-Rays from the Northwest Shell of the New
SNR RX J1713.7-3946:  The Second SN1006 ?}

\author{Katsuji {\sc Koyama},$^{1,6}$ Kenzo {\sc Kinugasa},$^2$
Keiichi {\sc Matsuzaki},$^3$ \\ Mamiko {\sc Nishiuchi},$^1$ 
Mutsumi {\sc Sugizaki},$^4$ Ken'ichi {\sc Torii}$^2$ \\
Shigeo {\sc Yamauchi},$^5$  and Bernd {\sc Aschenbach}$^7$ \\ [12pt]
 $^1$ {\it Department of Physics, Graduate School of Science,
Kyoto University, }\\ {\it Sakyo-ku, Kyoto 606-01}\\
{\it E-mail(KK): koyama@cr.scphys.kyoto-u.ac.jp}\\
 $^2$ {\it Department of Earth and Space Science, Faculty of Science, Osaka 
University, 1-1 Machikaneyama-cho,}\\ {\it Toyonaka, Osaka 560}\\
 $^3$ {\it Department of Physics, Graduate School of Science,
University of Tokyo,}\\ {\it Hongo, Bunkyo-ku, Tokyo}\\
 $^4$ {\it Institute of Space and Astronautical Science,
3-1-1 Yoshinodai, }\\ {\it Sagamihara, Kanagawa 229}\\
 $^5$ {\it Faculty of Humanities and Social Sciences, Iwate
University, 3-18-34 }\\ {\it Ueda, Morioka, Iwate 020}\\
 $^6$ {\it CREST: Japan Science and Technology Corporation (JST)}\\
 $^7$ {\it Max-Planck-Institut Fuer Extraterrestrische Physik,
 Giessenbachstrasse,}\\ {\it D-85740 Garching, Germany}}
\abst{
     We report ASCA results of a featureless X-ray spectrum from RX J1713.7$-$3946, 
a new shell-like SNR discovered with the ROSAT all sky survey.  The northwest 
part of RX J1713.7$-$3946 was in the field of the ASCA Galactic Plane Survey Project 
and was found to exhibit a shell-like structure.  The spectrum, however shows 
neither line emission nor any signature of a thermal origin.
Instead, a power-law model with a photon index 
of 2.4-2.5 gives reasonable fit to the spectrum, suggesting a non-thermal 
origin.  Together with the similarity to SN1006, we propose that RX J1713.7$-$3946 
is the second example, after SN1006, of a synchrotron X-ray radiation from a 
shell of SNRs.  Since the synchrotron X-rays suggest existence of extremely 
high energy charged particles in the SNR shell, our discovery should have 
strong impact on the origin of the cosmic X-rays.}

\kword{  ISM: individual objects (RX J1713.7$-$3946) --- ISM: supernova remnants ---Radiation mechanism ---X-rays: sources ---X-rays: spectra}

\maketitle
\thispagestyle{headings}

\section{Introduction}

X-rays from shell-like supernova remnants (SNRs) are 
generally attributable to a thin thermal plasma shock-heated by an energetic 
explosion of a progenitor, while additional hard X-rays with a power-law 
spectrum have been found  in some of the shell-like SNRs.  The most outstanding
and extreme example of a non-thermal emission from a shell is found in SN1006 
(Koyama et al. 1995).  The power-law X-ray spectrum extending smoothly to the 
radio band leads us to suspect that the X-rays are due to synchrotron 
radiation of extremely high energy electrons reaching to about 100 TeV 
(Koyama et al. 1995).  Similar, but somehow less compelling example for such 
synchrotron X-ray radiation is found in IC443; the hard X-ray image shows an 
arc structure surrounding a center-filled thermal emission 
(Keohane et al. 1997).  

These non-thermal emissions from SNR shells are 
potentially important with the conjecture of a possible origin of cosmic ray 
production and acceleration. The electro-magnetic radiation from high energy  
charged particles has a wide band spectrum from radio to gamma-rays, with  
radiation processes such as synchrotron, Bremsstrahlung, inverse-Compton 
scattering and nuclear reaction. However, such radiations from SNRs with no 
central pulsar have been scarcely observed, except for the synchrotron 
emission from electrons of energy below GeV (for radio SNRs). If the hard 
X-ray components found in SN1006 and IC443 really come from non-thermal 
electrons accelerated by a shock wave, protons should also be accelerated to 
the same energy, providing a strong support for the cosmic-ray acceleration  
in shell-like SNRs, possibly near to the "knee energy" (Reynolds 1996).  
Consequently, the most urgent subjects would be to find further evidence for a
non-thermal emission from other shell-like SNRs and to study whether the 
non-thermal emission is universal or not.  Previous non-imaging instruments or
soft X-ray imaging satellites have been unsuccessful for these studies.  A  
breakthrough may be provided by the wide X-ray band imaging spectroscopy, 
like ASCA, as was already demonstrated with two cases; SN1006 and IC 443. 

Pfeffermann and Aschenbach (1996; here and after PA96) 
discovered a shell-like SNR RX J1713.7$-$3946 with the ROSAT all sky survey in the 
Constellation Scorpius with the approximate center at R.A.$(2000)= 
17^{\rm h}13^{\rm m}\hspace{-5pt}.\hspace{2pt}42$,
Dec.$(2000)= -39^{\circ}\hspace{-5pt}.\hspace{2pt}46^{\prime} 
27^{\prime\prime}$  (hence named as RX J1713.7-3946).  The X-ray 
image shows a slightly elliptical of a maximum extent of 70$^{\prime}$, 
with an enhanced
emission at the northwest rim.  Thus they claimed this SNR as a shell-like 
SNR.  A composite type of shell-like plus synchrotron nebula around a putative
pulsar is also possible, because they found unidentified point-like sources  
at the center of the remnant.

This paper reports a peculiar X-ray spectrum from the 
bright northwest shell; unlike the shell-like appearance it shows an X-ray 
spectrum with no significant emission line.  Accordingly, we suggest this SNR 
to be another clear example, after SN1006, of a non-thermal emission from 
shell-like SNRs. 

\section{Observations}

The ASCA Galactic Plane Survey Project (here and after 
AGPSP) has started from the AO4 period, and is planed to cover all the 
galactic inner disk ($|l|$ $<$ 45$^{\circ}$ and $|b|$ $<$ 0$^{\circ}\hspace{-4.5pt}.
\hspace{.5pt}$8) with successive 
pointing observations of about 10 ksec exposure in a region of 50$^{\prime}$ 
diameter 
each, and will be continued during the full ASCA mission life of several 
years.  One major objective of the AGPSP is, utilizing the wide and high 
energy band ( up to 10 keV) X-ray imaging capability and high spectral 
resolving power of ASCA, to search for possible X-ray SNRs in the galactic 
inner disk, which would be either catalogued radio SNRs or completely new 
SNRs.    

The survey of the AO4 phase was made in 1996 
August-September, covered the region of $l=$342$^{\circ}\hspace{-4.5pt}.
\hspace{.5pt}2-346^{\circ}\hspace{-4.5pt}.\hspace{.5pt}$7, 
347$^{\circ}\hspace{-4.5pt}.\hspace{.5pt}2-351^{\circ}\hspace{-4.5pt}.\hspace{.5pt}$7 and 
$l$=1$^{\circ}\hspace{-4.5pt}.\hspace{.5pt}0-5^{\circ}
\hspace{-4.5pt}.\hspace{.5pt}$5 along 
$b$=0$^{\circ}$ of about 50$^{\prime}$ width. Some of the fields covered a 
fraction of the new SNR RX J1713.7$-$3946,  
the region of the bright northwest rim.  These observations were made with two
CCD cameras ( designated SIS0 and SIS1 ) and two gas-imaging spectrometers 
(GIS 2 and GIS 3) at foci of 4 thin-foil X-ray mirrors (XRT) on board the ASCA
satellite.   Details of the instruments are respectively given in 
Ohashi et al. (1996) and Makishima et al. (1996), Burke et al. (1994), and 
Serlemitsos et al. (1995),  for GISs, SISs and XRTs.   While, a general 
description of ASCA is given in Tanaka et al. (1994).  

The SIS data were obtained in the 4-CCD mode covering the 
field of view of 22$^{\prime}$ $\times$ 22$^{\prime}$ square,  
while the GIS data were taken  
in the normal PH mode  with a circular field of 25$^{\prime}$ radius.  
We excluded the data obtained at the South Atlantic Anomaly (SAA), the earth 
occultation, 
the high-background regions at low geomagnetic cut-off rigidities of $<$ 
6 GeV/c.  For the GIS data, we applied a rise-time discrimination technique 
to reject particle events.   The data with the elevation angle from the earth 
rim of $<$ 5$^{\circ}$ are also excluded in GIS, while for the SIS data 
screening criterion are $<$ 10$^{\circ}$ and $<$ 20$^{\circ}$(SIS1) and  
$<$ 25$^{\circ}$(SIS0) from the night and day earth rims, respectively.

The data quality, after about 3 years on orbit operation, 
were exhibited significant degradation due mainly to the particle 
irradiation (Dotani et al. 1995).  The most serious problem for the present 
SNR observation is the  increase of pixel-to-pixel fluctuation of the dark 
current, which makes the energy resolution worse.  In order to estimate and 
subtract the pixel-dependent dark current, we intentionally took the full 
pixel data (the frame mode data) during the normal operation, and estimated 
the dark frame level for the relevant CCD pixels.  This procedure, called the 
RDD (Residual Dark Distribution)-correction was found to restore  the
energy resolution significantly;  improved from ~210  eV to ~ 110 eV of 
FWHM at 1.86 keV in the 4 CCD mode.  Further details of the long term 
performance of the CCD in orbit and the RDD-correction technique are given by 
Dotani et al. (1995, 1997).

\section{Analysis and Results}

Figure 1 shows the GIS mosaic image of 4-successive 
pointings near the northwest part of RX J1713.7$-$3946.   In the figure, we see three
X-ray sources: point-like objects at northeast and southwest and a diffuse 
emission with shape of crescent reversed.  The first point source is a newly 
discovered transient and the other is 1RXS J170849.0-400910, from which we 
found a coherent pulsation of about 11-s (Sugizaki et al. 1997). Details of 
these source will be given in separate papers.   The center diffuse source 
comes near the ROSAT new SNR RX J1713.7$-$3946.  By comparing with the full image of 
RX J1713.7$-$3946  of ROSAT observation (PA96), we conclude that the diffuse 
structure is certainly a part of RX J1713.7$-$3946, the bright northwest rim.    
Although the SIS field of view is smaller than GIS, we also found a diffuse  
excess emission from the northwest shell. 

We extracted the GIS and SIS spectra from the shell-like 
region, and subtracted the nearby background sky. The northwest shell 
is located in the galactic ridge region, which exhibits an enhanced thin 
thermal emission, hence the background subtraction should be particularly  
careful.  Kaneda (1997) and Kaneda et al. (1997) reported the Galactic ridge 
X-rays observed with ASCA typically show two temperature components: the soft 
( about 0.8 keV) with scale height of about 3$^{\circ}$ and the hard (7 keV) 
with 
smaller scale height of 0$^{\circ}\hspace{-4.5pt}.\hspace{.5pt}$5.  
The center positions of the background 
regions we took, are nearly the same distance from the galactic plane as that
of the center of the shell.   Therefore the latitude dependence of the 
Galactic ridge emission would be canceled in the background subtraction.   
Furthermore the estimated background flux is about 6 $\times$ 10$^{-4}$  
counts~s$^{-1}$ arcmin$^{-2}$ GIS$^{-1}$ in the 0.7-10 keV band,    
which is 20 \%  of the mean surface density  of the northwest rim of 
3.2$\times$ 10$^{-3}$ counts~s$^{-1}$ arcmin$^{-2}$ GIS$^{-1}$.  
The spatial variation of the Galactic ridge emission found by Kaneda (1997)  
is less than a few 10 \% near the source position.  Therefore a systematic 
error of the background subtraction, attributable to possible uncertainty of 
the galactic diffuse emission, would be well below a few \% of the total 
source flux, hence we ignore this error.

Figure 2 shows the background subtracted SIS and GIS 
spectra.   Since no significant line is found in the spectra, we tried to set 
upper limits of the line equivalent widths (EWs) at 1.86, 2.46 and 6.7 keV, 
which respectively correspond to He-like Si, S and Fe lines and are expected 
to be major lines in a plasma of a few keV temperature.  The 90 \% upper limit
of the EWs of Si, S and Fe lines are respectively, 16, 35 and 82 eV for the 
GIS spectra, and 10, 12 and 186 eV for the SIS spectra.   These values are 
nearly to or less than 10 \% of those found in Cas A: 0.4-0.6, 0.4-0.5, 
0.6-1.3 keV of EW,  for Si, S and Fe, respectively (Holt et al. 1994).    
For a further investigation, we fitted the spectrum to a thin thermal model 
(MEKA model) with abundances varying collectively.   Although GIS spectra do 
not reject  (90 \% confidence),  the SIS spectra firmly reject the thin  
thermal model with more than 99.9 \% confidence level (reduced $\chi^{2}$ of 
1.45 for 179 d.o.f. ).   For a comparison and discussion, we list the best-fit 
parameter values in table 1.  Small differences of the best-fit parameters 
between the SIS and GIS spectra is mainly due to the difference of the regions
from which we have made the spectra; the larger area in GIS than in SIS, 
although possible calibration errors in GIS and SIS do exist.  The small 
difference, however, is  minor for further discussions.   From the table,  
we see that abundances are constrained to be unrealistically small; the 90 \% 
upper limit are  only 0.09 and 0.08 of solar for the GIS and SIS spectra,
respectively.  The plasma temperatures of 3.1 (SIS) - 3.8 (GIS) keV are higher
than any other shell-like SNRs; even  the youngest shell-like SNR Cas A and 
W49B exhibit significantly lower temperatures of about 2 keV 
( Holt et al. 1994, Fujimoto et al. 1995 ).  
	
Thus, together with the $\chi^{2}$-test and no emission 
line (or extremely low abundances), X-rays from the shell of RX J1713.7$-$3946 
can not be attributable to a thin thermal plasma. Accordingly we fit the 
spectra to a power-law with an interstellar absorption, then found to be 
acceptable with the best-fit parameters given in table 1.  The best-fit models
convolved with the detector responses are also shown in figure 2.

\section{Discussion}

PA96 divided the ROSAT data from RX J1713.7$-$3946 in several 
regions, and found that the spectra, on average, can be fitted with a thin 
thermal model of two cases; one is a high temperature plasma emission of  
averaged value of kT $=$ 4.8 keV, with position-dependent absorptions of  
$N\raisebox{-0.5ex}{\scriptsize H}$ $=$ (3-12) $\times$ 10$^{21}$ H~cm$^{-2}$, 
and the other is lower temperature plasma of kT $=$ 0.5 keV  with  
larger absorptions of $N\raisebox{-0.5ex}{\scriptsize H}$ $=$ (1.4-2) $\times$ 
10$^{22}$ H~cm$^{-2}$.   With the high temperature model,  
N\raisebox{-0.5ex}{\scriptsize H}  at the northwest rim is found to be  
(6.2$^{+}_{-}$1) $\times$ 10$^{21}$ H~cm$^{-2}$, in good agreement with a thin 
thermal fit of ASCA data  of $N\raisebox{-0.5ex}{\scriptsize H}$ $=$ (4.8$^{+}_{-}$ 0.4)$\times 10^{21}$  H~cm$^{-2}$ and  kT = 3.8$^{+}_{-}$ 0.3 keV.   
The ROSAT flux after removing the absorption for the high temperature 
solution, which is now found to be better approximation than that of lower 
temperature is F\raisebox{-0.5ex}{\scriptsize X}$= 4.4\times 10^{-10}$ 
erg ~s$^{-1}$cm$^{-2}$ in the whole SNR.  The ASCA flux within the 
northwest  rim is 2$\times 10^{-10}$ erg s$^{-1}$cm$^{-2}$  in the 0.5-10 keV 
band.  This value is  the same order of the total flux estimated with 
ROSAT, hence a significant fraction  of the X-rays from the SNR would be 
attributable to the bright northwest rim.  

The most important discovery is that the ASCA spectra, 
unlike the majority of shell like SNRs, have no significant emission 
line, and can  be fitted by a single power-law model.  This leads us to 
suspect that the X-rays from RX J1713.7$-$3946 are dominated by a non-thermal 
emission from the northwest shell. 

Non-thermal emissions have been found in many synchrotron 
nebulae (e.g. Kawai and Tamura 1996), in which high energy electrons 
are accelerated by a fast rotating neutron star. However in the northwest 
shell, we found neither point-like source nor any pulsed signal. These, 
together with the shell-like appearance, make the synchrotron nebulae scenario
unlikely to the northwest shell.   
Unclassified point sources are found near the SNR center with ROSAT, hence it
may still be  conceivable that a putative pulsar at the center emits 
relativistic jets, then  a synchrotron nebula located far from the pulsar can 
be made as is found in  the  SNR  W50 powered by the  unique source SS433 
(Yamauchi et al. 1994).   However, no pulsation is also found from the central
point sources.  Furthermore the morphology of W50 is, unlike to RX J1713.7$-$3946,  
not shell-like but plume-like shape.      	

More likely scenario is that RX J1713.7$-$3946 is the second 
case, after SN1006, of a shell-like SNR with a non-thermal dominant emission 
(Koyama et al. 1995).   Although the non-thermal X-ray flux is about 3 times 
larger than that from SN1006, no radio counterpart has been reported on this 
SNR. This may be partly due to the rather large background radio emission 
near the Galactic plane, but mainly due to the intrinsic faint radio flux.  
The faint radio emission indicates that the SNR should be in low density 
medium like the case of SN1006 lying off the plane.  Since RX J1713.7$-$3946 
is located  on the Galactic plane, the low density medium means that it is 
possibly lying in the intergalactic arm.   
	
The global similarity of RX J1713.7$-$3946 to SN1006 leads us to 
infer that essentially the same shock acceleration process as the 
SN1006 shell are  going on in the shell of this new SNR(Reynolds 1996). 
To investigate this scenario, in particular to determine the 
physical parameters such as the strength of the magnetic field, acceleration 
age, energy loss rate, and the diffusion time scale of the high energy 
particles, further observations,  not only in the X-ray domain but also the
radio and gamma ray bands, are strongly encouraged.  These are highly 
important in the conjecture of the long standing problem of the cosmic ray 
acceleration and its origin (see e.g. ,  Koyama et al. 1995, Reynolds 1996).   
	
The age estimation of PA96 based on the assumption of a thin
thermal plasma at Sedov phase is found now to be incorrect.  Thus 
the proposal  by Wang et al. (1997) that RX J1713.7$-$3946 is a remnant of a 
supernova of AD393 appeared in the old Chinese literature become more 
debatable than before.  Suppose that the ($N\raisebox{-0.5ex}{\scriptsize H}$ 
value of 10 kpc distance along the SNR  direction is 6$\times$10$^{22}$ 
H~cm$^{-2}$,  an averaged value to the Galactic center region, we infer that 
the SNR located at a distance of about 1 kpc, possibly just behind the 
Sgr-Carina arm.  This gives the radius of RX J1713.7$-$3946 (70$^{\prime}$ along the 
major axis) to be about 10 pc, the same size as SN1006. 
Taking the expansion speed of the SNR to be the largest so far found in young 
SNRs, or about 10$^{9}$ cm~s$^{-1}$, the age is given to be about 1000 
$yr$.   
This age should be taken as a lower limit, and the real age would be a few 
thousand years, still leaving a possibility of the AD393 SN origin.

\vspace{1pc}

We are grateful to all the member of ASCA team. Our  
particular thanks are due to the ASCA Galactic Plane Survey Team for  
fruitful collaborations and discussions. Thanks are due to R. Petre for his 
useful comments.

\vspace{1pc}

\section*{References}
\small

\re Burke B. E., Mountain R.W., Deniels P.J.,  Dolat V. S.  1994,  
	IEEE Trans. Nuclear Science  41, 375.
\re Dotani T., Yamashita A.,  Rasmussen A., the SIS team 1995,  
        ASCA News 3, 25
\re Dotani T., Yamashita A.,  Ezuka E., Takahashi K., Crew G., Mukai K.,  
        the SIS team 1997,  
	ASCA News 5, 14
\re Fujimoto R., Tanaka T., Inoue H., Ishida M., Itoh M., Mitsuda K., 
        Sonobe T., Tsunemi H. et al. 1995, 
	PASJ 47, L31.
\re Holt S.S.,  Gotthelf E.,  Tsunemi H., Negoro H. 1994,  PASJ  46,  L151
\re Kaneda N. 1997 Ph.D. thesis, University of Tokyo
\re Kaneda H., Makishima K., Yamauchi S., Koyama K., Matsuzaki K., 
        Yamasaki M.Y. et al. 1997,   	Ap J (submitted)
\re Kawai N.,  Tamura K. 1996, in X-ray Imaging and Spectroscopy of Cosmic 
        Hot Plasma, Universal Academy Press, Inc, Tokyo, Japan, eds F. 
        Makino and K. Mitsuda, p341 
\re Keohane. J.W.,  Petre R., Gotthelf E.V., Ozaki M., Koyama  K. 1997, 
	ApJ.  (accepted) 
\re Koyama. K, Petre R., Gotthelf E.V., Hwang U., Matsuura M., Ozaki M.,  
        Holt. S. 1995,  Nature  378,  255
\re Makishima K., Tashiro M., Ebisawa K., Ezawa H., Fukazawa Y., Gunji S., 
        Hirayama M., Idesawa E. et al. 1996,  PASJ  48, 171 
\re Mewe R., Gronenschild E.H.B.M., van den Oord G.H.J.  
        1985,  A\&AS  62, 197
\re Ohashi T., Ebisawa K., Fukazawa Y., Hiyoshi K., Horii M., Ikebe Y., 
        Ikeda H., Inoue H., et al. 1996,  PASJ  48, 157
\re Pfeffermann E., Aschenbach B., 1996, in Roentgenstrahlung from the 
        Universe, International Conference on X-ray Astronomy and 
        Astrophysics, MPE Report 263 (H.U. Zimmermann,  J.E. Truemper and H. 
        Yorke), P267  (PA96)
\re Reynolds S.P.   1996, ApJ  459,  L13
\re Serlemitsos P.J., Jolota L., Soong Y., Kunieda H., Tawara Y., 
        Tsusaka Y., Suzuki H., Sakima Y. et al. 1995,  PASJ  47, 105
\re Sugizaki  M., Nagase F., Yamauchi S., Koyama K., Matsuzaki K., 
        Asanuma T., Kinugasa K., Torii K. et al. 1997, IAU circular No 6585
\re Tanaka Y., Inoue H., Holt S.S. 1994,  PASJ  46,  L37
\re Yamauchi S., Kawai N., Aoki T. 1994,  PASJ 46,  L109
\re Wang Z.R., Qu -Y., Chen Y. 1997,  A\&A 318,  L59

\normalsize

\clearpage
\begin{table}[h]
 \begin{center}
         Table 1: The best-fit parameters of thin thermal  (MEKA) and  power-law models.\\
\begin{tabular}[]{lllll} \hline \hline
               & Temperature (kT)[keV] & $N\raisebox{-0.5ex}{\scriptsize H}$ [$10^{21}$cm$^{-2}$] & Abundance & r
educed-$\chi^{2}$(d.o.f) \\
\hline
GIS            & 3.8$^{+}_{-}$ 0.3 &   4.8$^{+}_{-}$ 0.4 &      $<$ 0.09  &         1.14  (144) \\
SIS            & 3.1$^{+}_{-}$ 0.2 &   7.4$^{+}_{-}$ 0.3 &      $<$ 0.08  &         1.45  (179) \\
\hline 
               &photon index($\alpha$) & $N\raisebox{-0.5ex}{\scriptsize H}$ [$10^{21}$cm$^{-2}$] &  & reduced-$\chi^{2}$(d.o.f) \\   
\hline
GIS            & 2.4$^{+}_{-}$ 0.1 &   8.1$^{+}_{-}$ 0.6  &                  &    1.12  (145) \\    
SIS            & 2.5$^{+}_{-}$ 0.1 &   10.1$^{+}_{-}$ 0.5 &                  &    1.23  (180) \\
\hline
\end{tabular}
\end{center}
\vspace{6pt}\par\noindent
$*$ Errors are single parameter 90\% confidence levels.\\
\par\noindent
$\dagger$ Assuming the solar abundance.\\
\end{table}

\section*{Figure Captions}
\small

Figure~1:A mosaic GIS image along the galactic plane near  ($l$, $b$)= 
           (347$^{\circ}$, 0$^{\circ}$).  
         The structure of crescent reversed  is a part of RX J1713.7$-$3946
         (the northwest rim), the other two point sources are respectively 
         a new transient source (the northeast source) and new X-ray pulsar 
         (the southwest source).

Figure~2:The SIS 0+1  and GIS 2+3  spectra of the northwest rim of 
         RX J1713.7$-$3946 (crosses) convolved with the detector response. 
         The histograms are the best-fit power law models (see text).  
         Lower panels for each are the data residuals from the best-fit 
         model in units of $\sigma$.

\clearpage

\begin{center}
  \epsfile{file=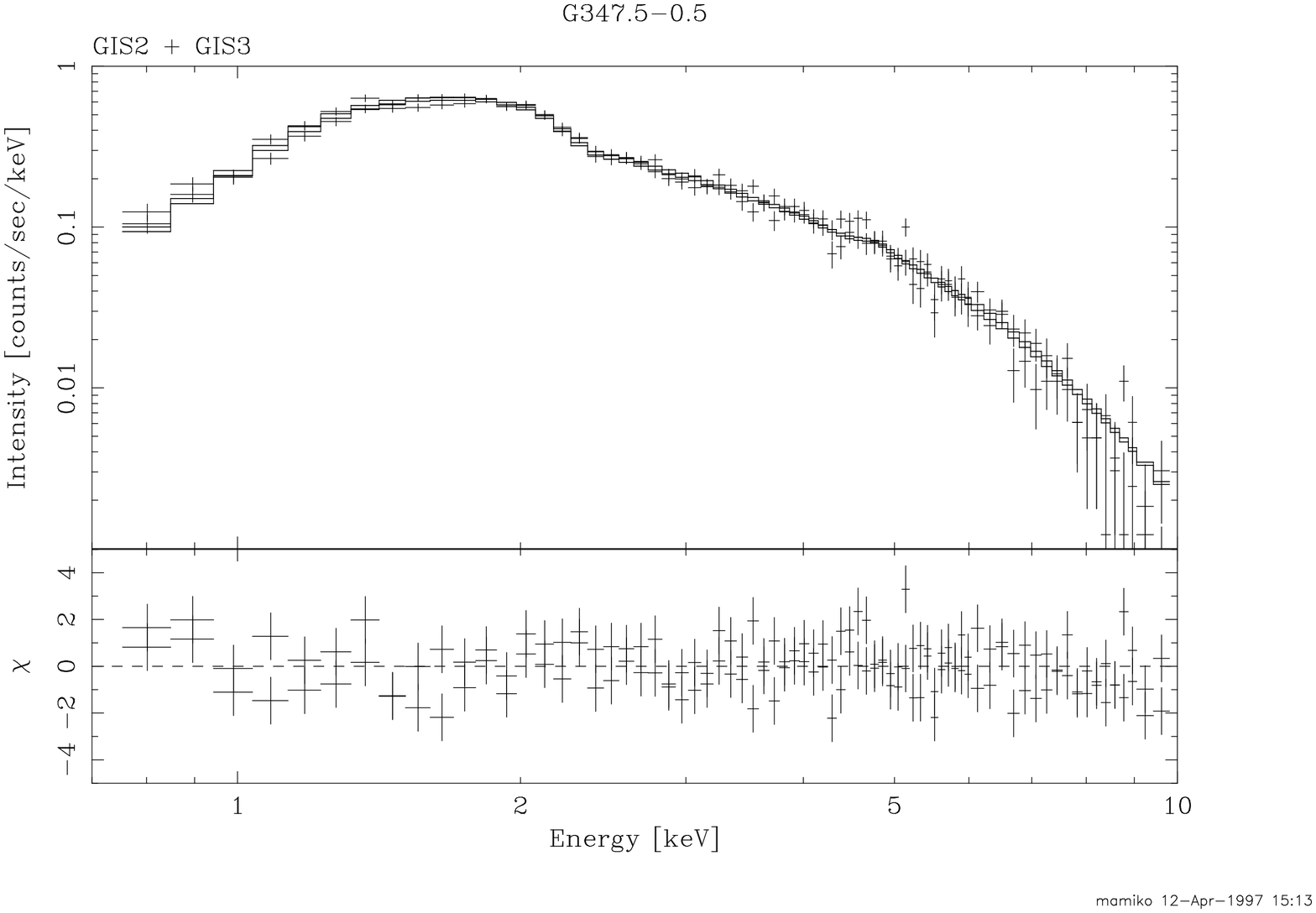,scale=0.5} \\
  \epsfile{file=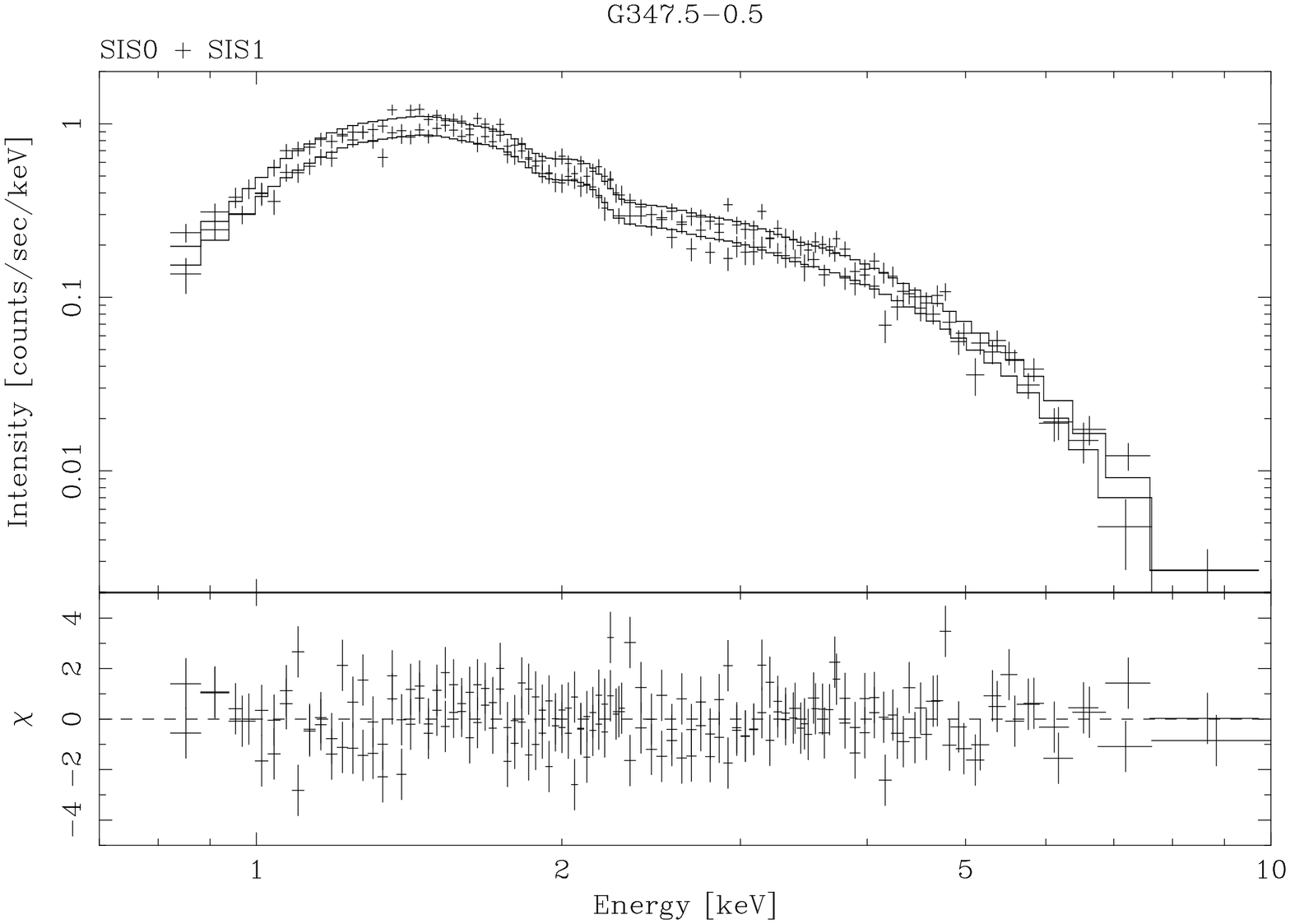,scale=0.5} \\
  Fugure~1: \\
\end{center}

\clearpage

\begin{center}
  \epsfile{file=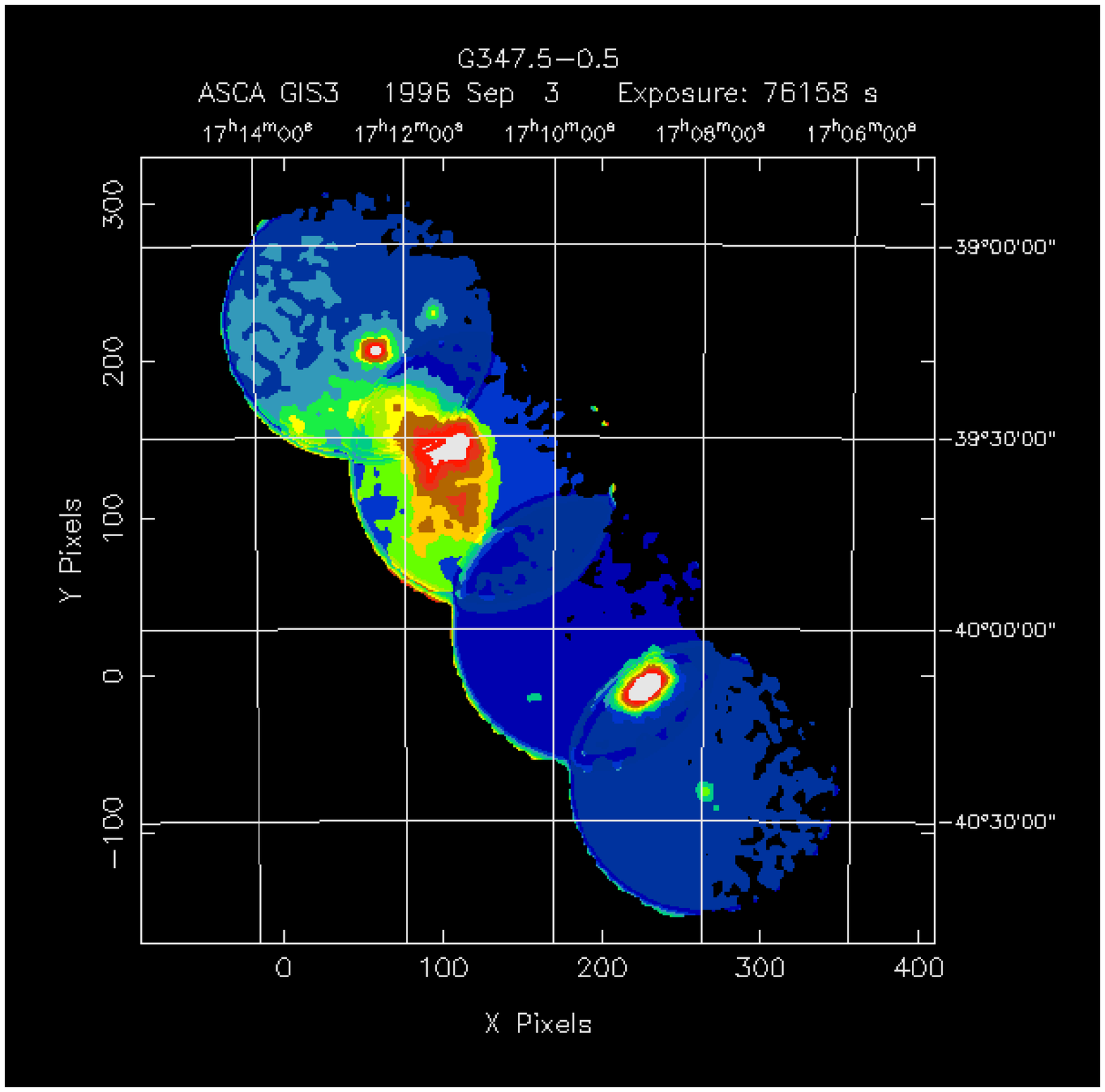} \\
  Figure~2: \\
\end{center}

\end{document}